\documentclass[12pt]{article}
\usepackage{latexsym}
\usepackage{amsmath}
\usepackage{amssymb}
\usepackage{amsthm}
\usepackage{graphicx}
\usepackage{subfigure}
\usepackage{hyperref}
\usepackage[latin1]{inputenc}
\usepackage[T1]{fontenc}
\usepackage[all]{xy}
\usepackage{lscape}
\makeatletter
\renewcommand{\@seccntformat}[1]
{\csname the#1\endcsname.\enspace} \makeatother
\def \XXint#1#2#3{{\setbox0=\hbox{$#1{#2#3}{\int}$}
     \vcenter{\hbox{$#2#3$}}\kern-.5\wd0}}

\newtheorem{example}{Example}
\newtheorem{definition}{Definition}
\newtheorem{proposition}{Proposition}
\newtheorem{algorithm}{Algorithm}

\begin{document}
\begin{center}
   {\bf An Optimal Multi-layer Reinsurance Policy under Conditional Tail Expectation}\\
{\sc  Amir T. Payandeh Najafabadi \footnote{Corresponding author:
amirtpayandeh@sbu.ac.ir; Phone no. +98-21-29903011; Fax no. +98-21-22431649} \& Ali Panahi Bazaz}\\
Department of Mathematical Sciences, Shahid Beheshti University,
G.C. Evin, 1983963113, Tehran, Iran.\\
\today
\end{center}
\begin{center}
    {\sc Abstract}
\end{center}
A usual reinsurance policy for insurance companies admits one or
two layers of the payment deductions. Under optimal criterion of
minimizing the conditional tail expectation (CTE) risk measure of
the insurer's total risk, this article generalized an optimal
stop-loss reinsurance policy to an optimal multi-layer reinsurance
policy. To achieve such optimal multi-layer reinsurance policy,
this article starts from a given optimal stop-loss reinsurance
policy $f(\cdot).$ In the first step, it cuts down an interval
$[0,\infty)$ into two intervals $[0,M_1)$ and $[M_1,\infty).$ By
shifting the origin of Cartesian coordinate system to
$(M_{1},f(M_{1})),$ and showing that under the $CTE$ criteria
$f(x)I_{[0, M_1)}(x)+(f(M_1)+f(x-M_1))I_{[M_1,\infty)}(x)$ is,
again, an optimal policy. This extension procedure can be repeated
to obtain an optimal k-layer reinsurance policy. Finally, unknown
parameters of the optimal multi-layer reinsurance policy are
estimated using some additional appropriate criteria. Three
simulation-based studies have been conducted to demonstrate: ({\bf
1}) The practical applications of our findings and ({\bf 2}) How
one may employ other appropriate criteria to estimate unknown
parameters of an optimal multi-layer contract. The multi-layer
reinsurance policy, similar to the original stop-loss reinsurance
policy is optimal, in a same sense. Moreover it has some other
optimal criteria which the original policy does not have. Under
optimal criterion of minimizing general translative and monotone
risk measure $\rho(\cdot)$ of {\it either} the insurer's total
risk {\it or} both the insurer's and the reinsurer's total risks,
this article (in its discussion) also extends a given optimal
reinsurance
contract $f(\cdot)$ to a multi-layer and continuous reinsurance policy.
\section{Introduction}
Designing an optimal reinsurance policy, in some sense, is one of
the most attractive aspects in actuarial science. Reinsurance is a
form of an insurance contract, that reinsurer accepts to pay a
portion of an insurer's risk by receiving a reinsurance premium.
Therefore, both reinsurance and insurance companies try to design
an optimal reinsurance policy to improve their ability to managing
their risks under a certain criteria, e.g., increasing their
surplus/wealth of company, decreasing the ruin probability, etc.

Several authors considered the problem of designing an optimal
reinsurance policy under a certain optimal criteria. Surprisingly,
in the most of studies the stop-loss reinsurance policy (or some
its modification) established as an optimal policy. For instance,
Borch (1960) proved that, under the variance retained risk optimal
criteria and in the class of reinsurance policies with an equal
reinsurance premium, the stop-loss reinsurance minimizes such
variance. Under Borch (1960)'s class of reinsurance policies ,
Hesselager (1990) showed that the stop-loss reinsurance is an
optimal policy which provides the smallest Lundberg's upper bound
for the ruin probability. Optimality of the one-layer stop-loss
contract under minimizes the ruin probability criteria and several
premium principles has been established by Kaluszka (2005).
Passalacqua (2007) studied impacts of multi-layer stop-loss
reinsurance contract on the valuation of risk capital (assessed
under the Solvency II framework) for credit insurance. Cai et al.
(2008) showed that the one-layer stop-loss contract is optimal
whenever {\it either} both the ceded and the retained loss
functions are increasing {\it or} the retained loss function is
increasing and left-continuous. Kaluszka \& Okolewski (2008)
established the one-layer stop-loss contract is an optimal
contract under the maximization of the expected utility, the
stability and the survival probability of the cedent. Tan et al.
(2011) and Chi \& Tan (2011) showed that under the expectation
premium principle assumption and the Conditional tail expectation
(CTE) minimization criteria the stop-loss reinsurance contract is
optimal. Porth et al. (2013) employed an empirical reinsurance
model (introduced by Weng, 2009) to show that, under the standard
deviation premium principle and consistency with market practice,
a one-layer stop-loss reinsurance contract is optimal. In a
situation that both the ceded and the retained loss functions are
constrained to be increasing and under the variance premium
principle assumption, Chi (2012) showed that one-layer stop-loss
reinsurance is always optimal over both the Value-at-Risk (VaR)
and the Conditional Value-at-Risk (CVaR) criteria. Ouyang \& Li
(2010) constructed a multi-layer reinsurance policy to achieve
sustainable development of an agricultural insurance policy in the
sense of adverse selection and mortal hazard problems. In 2012,
Dedu generalized the stop-loss reinsurance to a multi-layer
reinsurance policy. In the first step, she considered a certain
class of multi-layer reinsurance policies  with some unknown
parameters. An optimal reinsurance policy, in such class, have
been obtained by estimating unknown parameters such that the VaR
and the CTE of the insurer's total risk have been minimized. Chi
(2012) showed that under minimizes the risk-adjusted value of an
insurer's liability and the VaR (or the CVaR) criteria the
two-layer reinsurance contract is optimal under the Dutch premium
principle assumption. Cortes et al. (2013) considered a
multi-layer reinsurance contract consisting of a fixed number of
layers. Then, they determined an optimal multi-layer contract such
that for a given expected return the associated risk value is
minimized. Chi \& Tan (2013) established that a one-layer
stop-loss contract is always optimal over both the VaR and the
CVaR criteria and the prescribed premium principles. Cai \& Weng
(2014) showed under risk margin associated with an expectile risk
measure criteria a two-layer reinsurance contract minimizes the
liability of an insurer for a general class of reinsurance premium
principles. Panahi Bazaz \& Payandeh Najafabadi (2015) estimated
parameters of a one-layer reinsurance policy such that a convex
combination of the CTE of both the insurer's and reinsurer's
random risks are minimized. Optimality of the stop-loss contract
under distortion risk measures and premiums has been established
by Assa (2015). Zhuang et al. (2016) showed that in a situation
that the premium budget is not sufficiently high enough, under the
CVaR optimality criteria, the optimal reinsurance policy will
change from the stop-loss contract to a one-layer stop-loss.
Payandeh Najafabadi \& Panahi Bazaz (2016) considered a
co-reinsurance contract which is a combination of several
reinsurance contracts. Using a Bayesian approach parameters of
co-reinsurance contract have been estimated.

In order to exclude the moral hazard, an appropriate reinsurance
contract has to assign increasing functions to both insurer and
reinsurer portions. On the other hand, reported claims in
insurance industry have the property that higher claim size is
less frequent with more severe probability of loss. Whereas lower
claim sizes are  more frequent with less severe probability of
loss. Unfortunately, the stop-loss reinsurance contract despite
several well-known properties does not consider these two
important facts.

This article considers minimizing the $CTE$ risk measure of the
insurer's total risk as an optimal criterion to design an optimal
reinsurance contract. Then, it introduces an algorithm which
generalized a given optimal stop-loss policy to a multi-layer
optimal reinsurance policy. To achieve such optimal multi-layer
reinsurance policy, this article starts from a given optimal
stop-loss reinsurance policy $f(\cdot).$  In the first step, it
cuts down an interval $[0,\infty)$ into two intervals $[0,M_1)$
and $[M_1,\infty).$ By shifting the origin of Cartesian coordinate
system to $(M_{1},f(M_{1})),$ and showing that under the $CTE$
criteria $f(x)I_{[0,
M_1)}(x)+(f(M_1)+f(x-M_1))I_{[M_1,\infty)}(x)$ is, again, an
optimal policy. This extension procedure can be repeated to obtain
an optimal k-layer reinsurance policy. Finally, unknown parameters
of the multi-layer reinsurance policy are estimated using some
additional appropriate criteria. Practical application of our
findings have been shown through a simulation study. The
multi-layer reinsurance policy, similar to the original stop-loss
reinsurance policy is optimal, in a same sense. Moreover, it has
some other optimal criteria which the original policy does not
have. Under optimal criterion of minimizing a general translative
and monotone risk measure $\rho(\cdot)$ of {\it either} the
insurer's total risk {\it or} both the insurer's and the
reinsurer's total risks, this article (in its discussion) also
extends an optimal reinsurance contract $f(\cdot)$ to an optimal
multi-layer and continuous reinsurance policy.

This article is organized as the following. Section 2 collects
some elements that play vital roles in the rest of this article.
Moreover, Section 2 represents an algorithm that extends a given
optimal stop-loss reinsurance policy to an optimal multi-layer
policy. Section 3 describes three simulation-based studies
illustrating the practical application of our results. Parameters
of the optimal multi-layer contract, for each simulation study,
have been estimated using an additional appropriate criteria. In
Discussion results of this article (from two different senses)
extends an optimal reinsurance contract $f(\cdot),$ under a
general translative and monotone risk measure $\rho(\cdot),$ to an
optimal multi-layer and continuous reinsurance policy.
\section{Preliminary}
Suppose continuous and nonnegative random variable $X$ stands for
the aggregate claim initially assumed by an insurer. In addition,
suppose that random claim $X$ with a cumulative distribution
function $F_X(t)$ and a survival function $\bar F_X (t),$ and a
density function $f_X$ defines on the probability space $(\Omega,
\mathcal{F},P),$ where $\Omega=[0,\infty)$ and $\mathcal{F}$ is
the Borel $\sigma$-field on $\Omega.$  Now, let $X_I$ and $X_R,$
(or $X_R=h(X)$) respectively, stand for the insurer's and the
reinsurer's risk portions from random claim $X,$ such that
$X=X_I+X_R$ and $0 \le X_I~ \& ~X_R=h(X) \le X.$ Under this
presentation, the total risk of the insurance company can be
restated as
\begin{eqnarray}
\label{T_f}
\nonumber T_h(X)&=&X_I + \pi_h^X\\
       &=& X-h(X)+\pi_h^X,
\end{eqnarray}
where $h(\cdot)$ is a functional form of a reinsurance contract
and $\pi_h^X$ stands for a reinsurance premium.

Now, we collect some elements that play vital roles in the rest of
this article.
\begin{definition}
\label{translative-monotone-risk-measures} Risk measure
$\rho(\cdot)$ is called translative and monotone if and only if
$\rho(X+c)= \rho(X)+c$ and $\rho(X)\leq \rho(Y)$ whenever,
$P(X\leq Y)=1$ and $c\in{\Bbb R}.$
\end{definition}
In the sense of above definition a wide class of risk measures,
such as coherent, spectral, distortion, Quantile-based, and Wang,
are translative and monotone risk measures, see Denuit et al.
(2006) for other possible classes of translative and monotone risk
measures.

Consider the following class of reinsurance
policies.{
\begin{eqnarray}
\label{general-reinsurance-class}\nonumber \mathcal{C} &=& \left\{
h(X):~\hbox{both $h(X)$ and $X-h(X)$ are nondecreasing in
$X;$}\right.\\&&~~~~\left. 0 \le h(X)\le
X;~\hbox{and}~\pi_h^X=\hbox{constant}\right\},
\end{eqnarray}}\normalsize
where $\pi_h^X$ stands for the reinsurance premium under a
reinsurance contract $h(\cdot).$

Suppose $f(\cdot)$ in class of reinsurer contracts $\mathcal{C},$
given by \eqref{general-reinsurance-class}, minimizes given
translative and monotone risk measure $\rho(\cdot)$ of the total
risk of insurance company, i.e., $\displaystyle
f(X)\equiv\operatornamewithlimits{argmin}_{h\in\mathcal{C}}\rho\left(T_h(X)\right).$
Now one may cut down an interval $[0,\infty)$ into two intervals
$[0,M_1)$ and $[M_1,\infty)$ and shift the origin of Cartesian
coordinate system to $(M_{1},f(M_{1})),$ see Figure 1(a) for an
illustration. Again, in the new Cartesian coordinate system, the
shifted reinsurance contract $f(\cdot)$ is an optimal contract
and, in the old Cartesian coordinate system, the reinsurance
contract $g(x)=f(x)I_{[0,
M_1)}(x)+(f(M_1)+f(x-M_1))I_{[M_1,\infty)}(x)$ is an appropriate
contract. Since $f(\cdot)$ is an optimal contract, optimality of
$g(\cdot)$ arrives by showing that $\rho(T_g(X))\equiv
\rho(T_f(X)).$ Unfortunately proof of such identity is not
available for general translative and monotone risk measures.
Hopefully, Tan et al. (2011, Theorem 3.1) showed that under the
$CTE$ criteria as far as $g(\cdot)\in\mathcal{C}$ and $0\leq
g(x)\leq f^*(x)=\max\{x-d_\alpha, 0\},$ for a given
$\alpha\in(0,1)$ and all $x\geq0,$ any contract $g(\cdot)$ is
again optimal, i.e., $\rho(T_g(X))\equiv \rho(T_{f^*}(X)).$ Using
such seminal result, we can conclude that under the $CTE$
minimization criteria, the new contract $g(x)=f^*(x)I_{[0,
M_1)}(x)+(f^*(M_1)+f^*(x-M_1))I_{[M_1,\infty)}(x)$ is optimal.
Again cutting down an interval $[M_1,\infty)$ into two intervals
$[M_1, M_2)$ and $[M_2,\infty)$ and shifting the origin of
Cartesian coordinate system to $(M_{2},f^*(M_{2}-M_1)),$ we can
obtain new contract $f^*(x)I_{[0,
M_1)}(x)+(f^*(M_1)+f^*(x-M_1))I_{[M_1,M_2)}(x)+(f^*(M_2)+f^*(x-M_2))I_{[M_2,\infty)}(x)$
which Tan et al. (2011, Theorem 3.1) warranties its optimality.
Several implementation of the above procedure leads to an optimal
multi-layer reinsurance contract, under the $CTE$ minimization
criteria. The following algorithm provides such multi-layer
contract.
\begin{algorithm}
\label{algorithm-multi-layer-reinsurance-policy} Suppose $X_R$
stands for the reinsurer's risk portion from random claim $X.$ The
following steps design a multi-layer reinsurance policy which
minimizes the $CTE$ of the insurer's total risk.
\begin{description}
    \item[Step 1)] A multi-layer reinsurance policy is
    obtained by the following iterative algorithm:
    \begin{description}
        \item[Part 1)] For $k\geq2;$ Cut down an interval $[M_k,\infty)$ into two intervals
$[M_k, M_{k+1})$ and $[M_{k+1},\infty)$ and define the reinsurer's
risk portion by
\begin{eqnarray} \label{X_R_Algorithm_CTE}
  f_{k}(X) &=& f_{k-1}(X)I_{[0,M_{k})}(X)+\left[f_{k-1}(M_{k})+f(X-M_{k})
\right]I_{[M_{k},\infty)}(X),
\end{eqnarray}
where $f_0(X)=f(X)=\max\{X-d_\alpha,0 \};$
        \item[Part 2)] Go to Step 2 if a given stop criteria is met, otherwise set $k=k+1$ and go
        to Part (1)
    \end{description}
    \item[Step 2)]
        \begin{description}
        \item[Part 1)] The reinsurer's risk portion under the k-layer reinsurance policy is $X_R=f(X)I_{[0,M_1)}(X)+\sum_{j=1}^{k-1}f_j(X)I_{[M_{j},M_{j+1})}(X)+\left[f_{k-1}(M_{k})+f(X-M_{k}) \right]I_{[M_{k},\infty)}(X).$
        \item[Part 2)] Now estimate unknown parameters by some additional appropriate criteria (or estimation methods) along the fact that the fact that $E(\max\{X-d_\alpha,0\})=E(X_R).$
    \end{description}
\end{description}
\end{algorithm}
Closeness to an appropriate criteria (such as an optimal ruin
probability) can be considered, in advance, as a stopping criteria
in the above algorithm.

Algorithm \eqref{algorithm-multi-layer-reinsurance-policy} designs
an optimal multi-layer reinsurance policy which the insurer's and
the reinsurer's portion of both companies are increasing functions
in the initial insurer claim $X.$ Moreover it provides a sharing
system that its higher layer works appropriately for large
reported claim size.

Application of Algorithm
\eqref{algorithm-multi-layer-reinsurance-policy} leads to the
following optimal k-layer reinsurance policy. \footnotesize{
\begin{eqnarray}\label{X_R_k_layer_stop-loss-equal-unequal}
X_R^{opt} = \left\{ {\begin{array}{*{20}{c}}
    0&{X < {d_\alpha }}\\
    {X - {d_\alpha }}&{{d_\alpha } \le X < {M_1}}\\
    {{M_1} - {d_\alpha }}&{{M_1} \le X < {M_1} + {d_\alpha }}\\
    {X - 2{d_\alpha }}&{{M_1} + {d_\alpha } \le X < {M_2}}\\
    \vdots &{}\\
    {{M_{k}} - k{d_\alpha }}&{{M_{k}} \le X < {M_{k}} + {d_\alpha }}\\
    {X - k{d_\alpha }}&{{M_{k}} + {d_\alpha } \le X}
    \end{array}} \right.
\end{eqnarray}}\normalsize
Figure 1(b) illustrate optimal multi-layer reinsurance policy
\eqref{X_R_k_layer_stop-loss-equal-unequal}.
\begin{center}
\begin{figure}[h!]
\centering\subfigure[]{
\includegraphics[width=7cm,height=6cm]{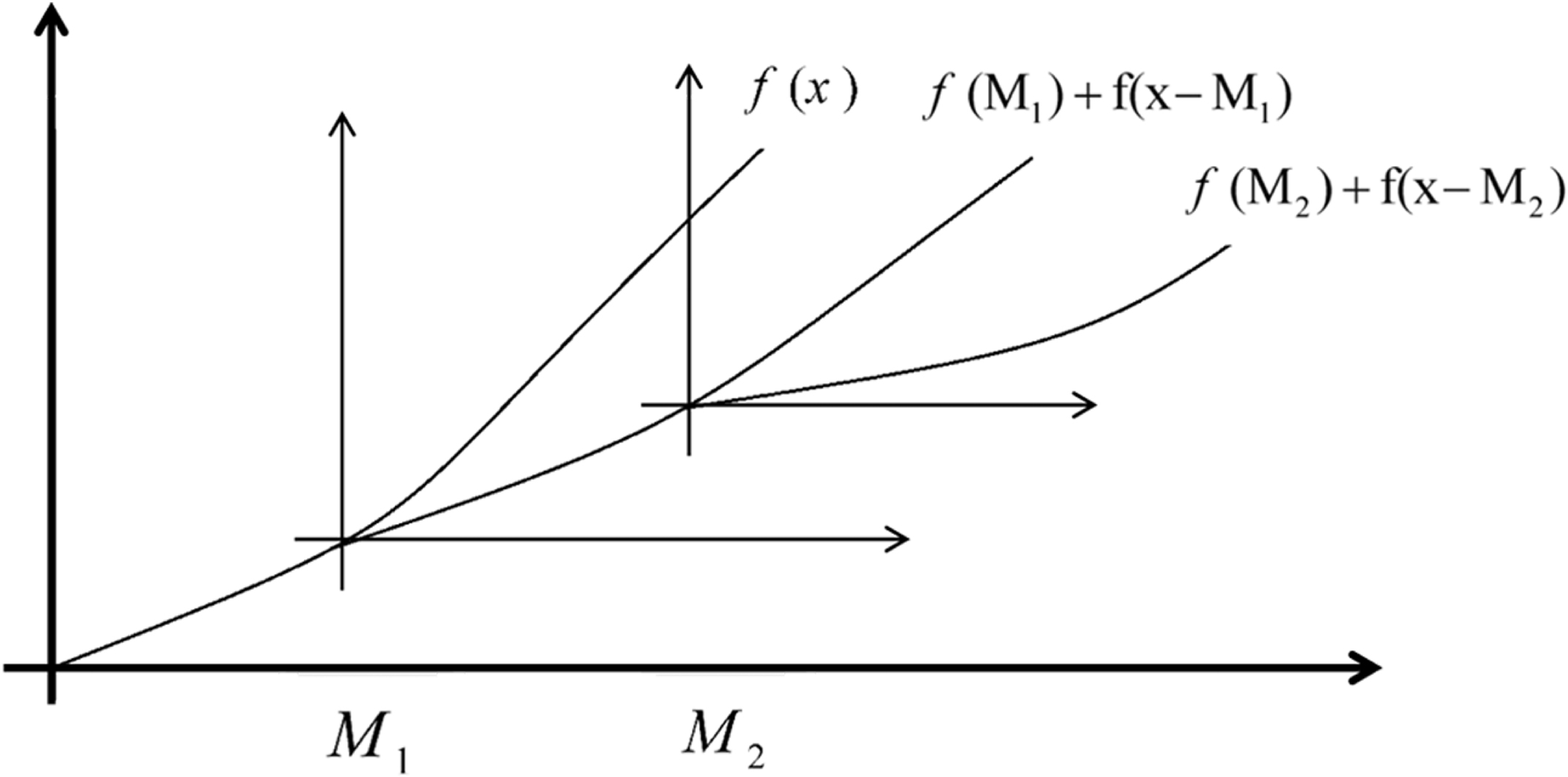}}\subfigure[]{
\includegraphics[width=7cm,height=7cm]{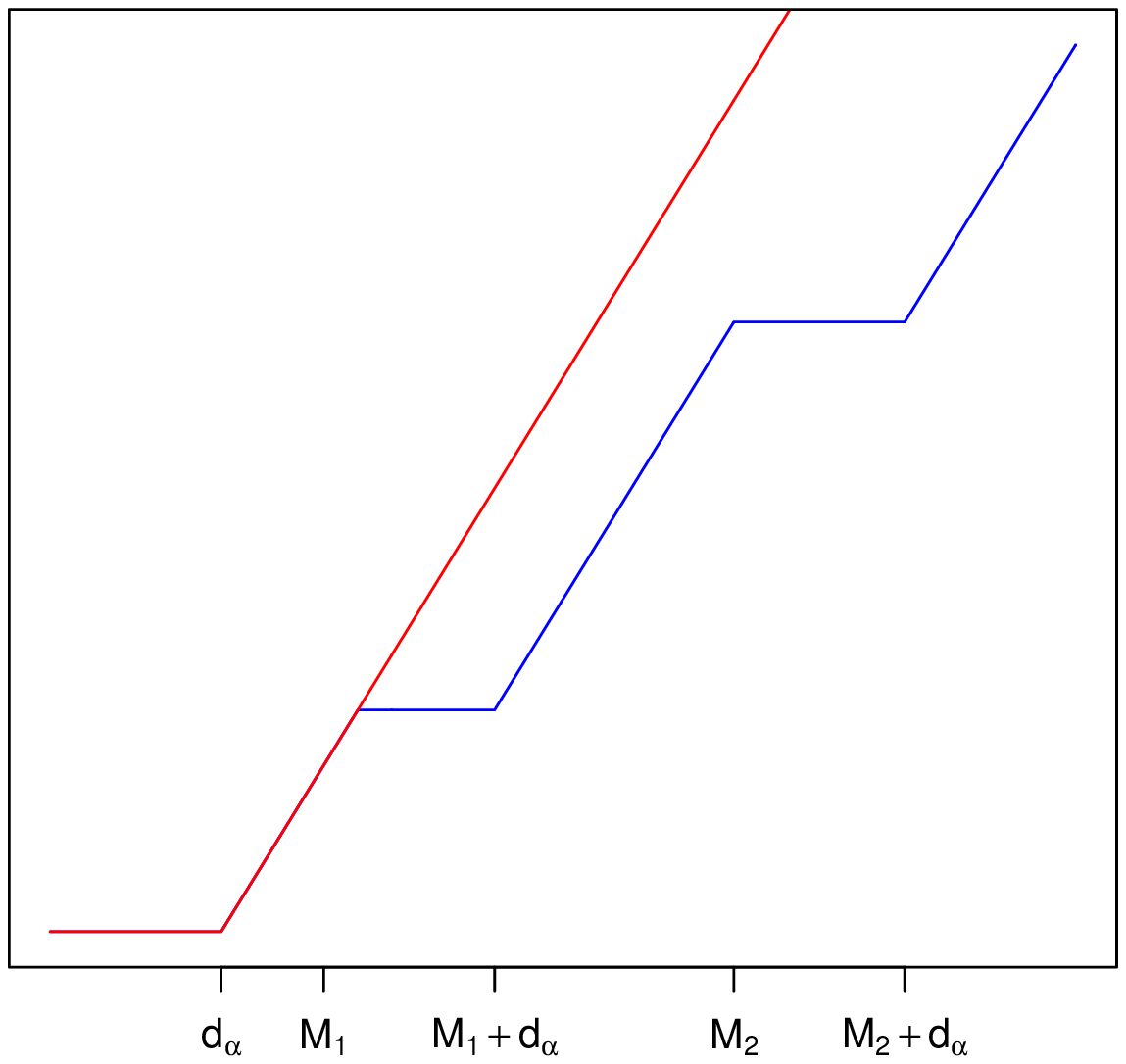}}
\caption{\scriptsize Part (a): Shifting the Cartesian coordinate
system and finding the optimal contract in the new Cartesian
coordinate system and Part (b): Stop-loss and an optimal and
k-layer reinsurance strategies.}
\end{figure}
\end{center}
For the sake of simplicity, hereafter now, we set
$M_0^*:=d_\alpha,$ $M_1^*:=M_1,$ $M_2^*:=M_1+d_\alpha$ and so on.

The cumulative distribution function for optimal k-layer
reinsurance policy \eqref{X_R_k_layer_stop-loss-equal-unequal} can
be restated as {\footnotesize
\begin{eqnarray*}
  {F_{X^{opt}_R}}(t) &=& {F_X}\left(t_{-}+M^*_0\right)I_{[0, M^*_1-M^*_0)}(t)+{F_X}\left(t+M^*_2- (M^*_1-M^*_0)\right)I_{[M^*_1-M^*_0, (M^*_3-M^*_2)+ (M^*_1-M^*_0))}(t) \\
  &&+{F_X}\left(t+M^*_4- (M^*_3-M^*_2)- (M^*_1-M^*_0)\right)I_{[(M^*_3-M^*_2)+ (M^*_1-M^*_0), (M^*_5-M^*_4)+(M^*_3-M^*_2)+ (M^*_1-M^*_0))}(t)\\
  &&+\cdots+{F_X}\left(t+ M^*_{m-2}-\sum\limits_{j = 1}^{k/2-2}{({M^*_{2j+1}} - {M^*_{2j }})}- (M^*_1-M^*_0)\right)I_{[\sum\limits_{j = 1}^{k/2-2}{({M^*_{2j+1}} -{M^*_{2j }})}+ (M^*_1-M^*_0), \infty)}(t);
\end{eqnarray*}}\normalsize

The following provides the moment generating function for the
reinsurer's risk portion from random claim $X,$ under optimal
k-layer reinsurance policy
\eqref{X_R_k_layer_stop-loss-equal-unequal}.
\begin{proposition}
    \label{moment_generating_X_R} Suppose $X_R$ stands for the reinsurer's risk portion from random claim $X,$ under
an optimal k-layer reinsurance policy which minimizes the $CTE$ of
the insurer's total risk. Then, the moment generating function for
the reinsurer's risk portion
    $X_R^{opt}$ under an optimal k-layer reinsurance policy.
    \begin{eqnarray*}
        {M_{X_R^{opt}}}(t) = && 1 - {e^{t(  ({M^*_1} - {M^*_0}) + \sum\limits_{j = 1}^{k/2- 1} {({M^*_{2j + 1}} - {M^*_{2j}})} )}}\bar F_X({M^*_{k - 2}}) + \int_{{M^*_0}}^{{M^*_1}} {t  {e^{t  (X - {M^*_0})}}\bar F_X(x)dx} \\
        &&+ \sum\limits_{j = 1}^{k/2- 1} {\int_{{M^*_{2j}}}^{{M^*_{2j + 1}}} {t{e^{t(x +   ({M^*_1} - {M^*_0}) + \sum\limits_{i = 1}^{j - 1} {({M^*_{2i + 1}} - {M^*_{2i}}) - {M^*_{2j}}} )}}\bar F_X(x)dx} } \\
        &&+ \int_{{M^*_{k - 2}}}^\infty  {{e^{t(X + \sum\limits_{j = 1}^{k/2- 1} {({M^*_{2j + 1}} - {M^*_{2j}})}  +   ({M^*_1} - {M^*_0}) - {M^*_{k -
        2}})}}dF_X(x)},
    \end{eqnarray*}
where $\sum_{j=a}^{b}c_j=0$ whenever $b<a.$
\end{proposition}
{\it Proof.} Observe that the moment generating function of
$X_R^{opt},$ given by Equation
\eqref{X_R_k_layer_stop-loss-equal-unequal} can be calculated as
follows
\begin{eqnarray*}
    {M_{X_R^{opt}}}(t) = &&\int_0^{{M^*_0}} {dF_X(x)}  + \int_{{M^*_0}}^{{M^*_1}} {{e^{t  (x - {M^*_0})}}dF_X(x)}  +  \cdots + \int_{{M^*_{m - 2}}}^\infty  {{e^{t(X + \sum\limits_{j = 1}^{k/2-1} {({M^*_{2j + 1}} - {M^*_{2j}})}  +   ({M^*_1} - {M^*_0}) - {M^*_{k - 2}})}}}
\end{eqnarray*}
The odd terms can be evaluated directly. The following calculation
represents that how one cab evaluate other terms.
\begin{eqnarray*}
    \int_{{M^*_2}}^{{M^*_3}} {{e^{t(x +   ({M^*_1} - {M^*_0}) - {M^*_2})}}} dF_X(x) &=& {e^{t(x +   ({M^*_1} - {M^*_0}) - {M^*_2})}}F_X(x)\left| {_{{M^*_2}}^{{M^*_3}}} \right. - \int_{{M^*_2}}^{{M^*_3}} {t{e^{t(x +   ({M^*_1} - {M^*_0}) - {M^*_2})}}F_X(x)dx} \\
    &=& {e^{t({M^*_3} +   ({M^*_1} - {M^*_0}) - {M^*_2})}}F_X({M^*_3}) - {e^{t({M^*_2} +   ({M^*_1} - {M^*_0}) - {M^*_2})}}F_X({M^*_2})\\
    &&- \int_{{M^*_2}}^{{M^*_3}} {t{e^{t(x +   ({M^*_1} - {M^*_0}) - {M^*_4})}}F_X(x)dx} \\
    &=& {e^{t({M^*_3} +   ({M^*_1} - {M^*_0}) - {M^*_2})}}F_X({M^*_3}) - {e^{t  ({M^*_1} - {M^*_0})}}F_X({M^*_2})\\
    &&- \int_{{M^*_2}}^{{M^*_3}} {t{e^{t(x +   ({M^*_1} - {M^*_0}) - {M^*_2})}}dx}  + \int_{{M^*_2}}^{{M^*_3}} {t{e^{t(x +   ({M^*_1} - {M^*_0}) - {M^*_2})}}\bar F_X(x)dx} \\
    &=& {e^{t({M^*_3} +   ({M^*_1} - {M^*_0}) - {M^*_2})}}F_X({M^*_3}) - {e^{t  ({M^*_1} - {M^*_0})}}F_X({M^*_2})\\
    &&- {e^{t({M^*_3} +   ({M^*_1} - {M^*_0}) - {M^*_2})}} + {e^{t  ({M^*_1} - {M^*_0})}} + \int_{{M^*_2}}^{{M^*_3}} {t{e^{t(x +   ({M^*_1} - {M^*_0}) - {M^*_2})}}\bar
    F_X(x)dx}.
\end{eqnarray*}
The desired proof arrives by a straightforward calculation.
$\square$

Similar to Proposition \eqref{moment_generating_X_R}, one may show
that under the optimal k-layer reinsurance contract, the moment
generating function for the insurer's risk portion,
$X_I=X-X_R^{opt},$ from random claim $X,$ is
\begin{eqnarray*}
        {M_{X - X_R^{opt}}}(t) &=&\bar F_X(0) - {e^{t({M_{k - 2}} - \sum\limits_{j = 1}^{k/2-2} {({M_{2j + 1}} - {M_{2j}})}  -   ({M_1} - {M_0}))}}\bar F_X({M_{k - 2}}) + \int_0^{{M_0}} {t{e^{tx}}\bar F_X(x)dx} \\
        &&+ \sum\limits_{j = 1}^{k/2-1} {\int_{{M_{2j - 1}}}^{{M_{2j}}} {t{e^{t(x -   ({M_1} - {M_0}) - \sum\limits_{i = 1}^{j - 1} {({M_{2i + 1}} - {M_{2i}})} )}}\bar F_X(x)dx} } \\
        &&+ {e^{t({M_{k - 2}} -   ({M_1} - {M_0}) - \sum\limits_{j = 1}^{k/2-2} {({M_{2j + 1}} - {M_{2j}})} )}}\bar F_X({M_{k -
        2}}),
    \end{eqnarray*}
where $\sum_{j=a}^{b}c_j=0$ whenever $b<a.$

Using Proposition \eqref{moment_generating_X_R} the expectation of
the reinsurer's risk portion $X_R^{opt},$ under an optimal k-layer
reinsurance can be evaluated as {\footnotesize
\begin{eqnarray*}
    E(X_R^{opt}) &=&{M^*_0}(F_X({M^*_1}) - F_X({M^*_0})) + \int_{{M^*_0}}^{{M^*_1}} {\bar F_X(x)dx}  + \sum\limits_{j = 1}^{k/2-2} {\int_{2j}^{2j + 1} {\bar F_X(x)dx} }\\
    &&+ \int_{{M^*_{k - 2}}}^\infty  {xdF_X(x)}  - {M^*_{k- 2}}(1 -
    F_X({M^*_{k-2}})).
\end{eqnarray*}}\normalsize
The next section conducts several simulation-based studies, to
show ``how one can employ some other appropriate criteria to fully
determine an optimal k-layer reinsurance contract''.
\section{Simulation Study}

This section provides four numerical examples to show how the
above findings, along with some other additional appropriate
criteria , can be applied in practice. These examples consider a
given multi-layer reinsurance policies which arrives by an
extension of the optimal stop-loss reinsurance policy. Unknown
parameters of each multi-layer reinsurance policy are estimated
using an additional appropriate criteria.

Borch (1960) showed that, under the variance retained risk optimal
criteria, in class of reinsurance contracts $\mathcal{C},$ given
by Equation \eqref{general-reinsurance-class}, the stop-loss
reinsurance is optimal. The following shows that the proportional
reinsurance contract minimizes a convex combination of variance of
the insurer's and the reinsurer's risk portions from random claim
$X.$

\begin{proposition}\label{minimal_combin_Var}
Suppose $X_R=h(X)$ and $X_I=X-h(X)$, respectively, stand for the
reinsurer's and the insurer's risk portions from random claim $X.$
Then, in class of reinsurance contracts $\mathcal{C},$ given by
Equation \eqref{general-reinsurance-class}, proportional contract
$h^*(X) = \frac{1}{{1 + \omega }}X$ minimizes the following convex
combination of variance of $X_R=h(X)$ and $X_I=X-h(X)$
$${Q_h} =\omega Var(h(X)) + (1 - \omega )Var(X - h(X)),$$ where
$\omega\in[0,1].$
\end{proposition}
{\it Proof.} The above convex combination of two variances can be
restated as{\footnotesize
   \begin{eqnarray*}
    \operatornamewithlimits{argmin}_{h\in\mathcal{C}}{Q_h} &=&\operatornamewithlimits{argmin}_{h\in\mathcal{C}}\left\{ \omega Var(h(X) - X + X) + (1 - \omega )Var(X - h(X))\right\}\\
    &=&\operatornamewithlimits{argmin}_{h\in\mathcal{C}}\left\{ \omega Var(X - (X - h(x))) + (1 - \omega )Var(X - h(X))\right\}\\
    &=&\operatornamewithlimits{argmin}_{h\in\mathcal{C}}\left\{ \omega Var(X) + Var(X - h(X)) - 2\omega Cov(X,X - h(X))\right\}\\
    &=&\operatornamewithlimits{argmin}_{h\in\mathcal{C}}\left\{ Var(X - h(X)) - 2\omega Cov(X,X - h(X))\right\}\\
    &=&\operatornamewithlimits{argmin}_{h\in\mathcal{C}}\left\{ E[{(X - h(X))^2}] - {[E(X - h(X))]^2} - 2\omega E[(X - h(X))X] + 2\omega E[(X - h(X))]E[X]\right\}\\
    &=&\operatornamewithlimits{argmin}_{h\in\mathcal{C}}\left\{ E\left[{(X - h(X))^2} - 2\omega (X - h(X))X\right] - E[(X - h(X))]\left[E[(X - h(X))] - 2\omega E(X)\right]\right\}\\
    &=&\operatornamewithlimits{argmin}_{h\in\mathcal{C}}\left\{ E[[(X - h(X))][(X - h(X)) - 2\omega X]] - E[(X - h(X))]E[(1 - 2\omega )X - h(X)]\right\}\\
    &=&\operatornamewithlimits{argmin}_{h\in\mathcal{C}}\left\{ Cov[(X - h(X)),(1 - 2\omega )X -
    h(X)]\right\}.
    \end{eqnarray*}}\normalsize
Therefore, one may conclude that the above convex combination is
minimal whenever $(X - h(X))$ and $(1 - 2\omega )X - h(X)]$ are
linearly dependent. Choosing $(1 - 2\omega )X - h(X) = \beta_0  +
\beta_1(X - h(X))$ leads to $h(X) =(1-2\omega-\beta_1)X/(1 -
\beta_1)-\beta_0/(1 - \beta_1).$ The fact that $0\leq h(X)\leq X$
implies that $\beta_0=0.$ Now by substituting back $h(X)
=(1-2\omega-\beta_1)X/(1 - \beta_1)$ in the above convex
combination, we have
$${Q_I} =[\omega \frac{{{{(1 - 2\omega  - \beta_1 )}^2}}}{{{{(1 - \beta_1 )}^2}}} + (1 - \omega )\frac{{{{(2\omega )}^2}}}{{{{(1 - \beta_1
)}^2}}}]Var(X).$$ Minimizing this expression, with respect to
$\beta_1,$ leads to desired result. $\square$

Proposition \eqref{minimal_combin_Var} shows that the proportional
reinsurance the contract minimizes a convex combination of
variance of $X_R$ and $X-X_R.$ The following example considers
this observation as an appropriate criteria to estimate unknown
parameters of an optimal 2-layer contract.

\begin{example}\label{Variance_criteria}
Suppose that random claim $X$ has been distributed according to
one of the distributions given in the first column of Table 1.
Moreover suppose that the optimal multi-layer contract has 2
layers and restated as
    \begin{eqnarray*}
        {X_R^{2-layer}}(X) = \left\{ \begin{array}{l}
            0\,\,\,\,\,\,\,\,\,\,\,\,\,\,\,\,\,\,\,\,\,\,\,\,\,\,\,\,\,\,\,\,\,\,\,\,\,\,\,\,\,\,\,\,\,\,\,\,\,\,\,\,\,\,\,\,\,\,\,\,\,\,\,\,\,\,\,X < {d_\alpha}\\
            X -{d_\alpha} \,\,\,\,\,\,\,\,\,\,\,\,\,\,\,\,\,\,\,\,\,\,\,\,\,\,\,\,\,\,\,{d_\alpha} \le X < {M_1}\\
            {M_1} - {d_\alpha} \,\,\,\,\,\,\,\,\,\,\,\,\,\,\,\,\,\,\,\,\,\,\,\,\,\,\,\,\,\,{M_1} \le X < {M_1}+{d_\alpha}\\
            X - 2{d_\alpha}\,\,\,\,\,\,\,{M_1}+{d_\alpha} \le X < M_2\\
            M_2-2{d_\alpha}\,\,\,\,\,\,\,{M_2} \le X < M_2+{d_\alpha}\\
            X - 3{d_\alpha}\,\,\,\,\,\,\,{M_2}+{d_\alpha} \le X \\
        \end{array} \right.
    \end{eqnarray*}
For the sake of simplicity, we set $M_1=d_\alpha+ d_1$ and
$M_2=2d_\alpha+d_1+d_2.$ Now $M_0$ has been estimated such that
$E(X_R)=E(\max\{X-d_\alpha, 0\}).$ Other two parameters $d_1$ and
$d_2$ have been estimated such that the square distance
$\left[Q_{{X_R^{2-layer}}}-Q_{h^*}\right]^2$ is minimized, where
$Q_{h}$ and $h^*$ are given in Proposition
\eqref{minimal_combin_Var}.
\end{example}
Table 1 shows estimation for unknown parameters of the above
optimal 2-layer ${X_R^{2-layer}}.$
\begin{center}
    \tiny{Table 1: Estimation for unknown parameters of the optimal
2-layer contract under variance optimal criteria, whenever $\omega=0.2$ and $\alpha=0.1$.\\
        \begin{tabular}[c]{c c c c c c c c  c c c}
            \hline
            Random claim distribution& $d_\alpha$ &$M_1$&$M_2$&$E(h^{SL}(X))=E(h^{2-layer}(X))$ & $CTE_{h^{SL}}=CTE_{h^{2-layer}}$&$Q_{h^*}$&$Q_{h^{SL}}$&$Q_{h^{2-layer}}$\\
            \hline
            Exp(10)       &23.0259&24.4258&48.4516&$1$ &10.423&16.667&52.948&46.1586\\
            Exp(8)        &18.4206&26.4986&45.9192&$0.4498$ &8.14&6.6707&33.8867&29.5415\\
            Exp(4)        &9.2103&13.2103&18.1928&$0.4099$ &4.0743&1.6692&8.4717&7.3853\\
            Weibull(1,2)  &1.5174&4.1396&6.657&$0.028$ & 0.2865&0.0358&0.1639&0.1338\\
            Weibull(3,2)  &4.5523&12.7469&18.2992&$0.02135$ &1.2235&0.322&1.475&1.204\\
            \hline
        \end{tabular}}\\
            $Q_{h}$ and $h^*$ are given in Proposition \eqref{minimal_combin_Var}, $h^{SL}(X)=\max\{X-d_\alpha,0\}$ and $h^{2-layer}(X)=X_R^{2-layer}(X).$ \\
        \normalsize
    \end{center}
The last three columns of Table 1 show the convex combination of
variance of $X_R=h(X)$ and $X_I=X-h(X)$ for optimal stop-loss,
optimal 2-layer and proportional (given by Proposition,
\ref{minimal_combin_Var}) contracts, respectively. As one may
observe that, under the optimal 2-layer contract such convex
combination of variances, compare to optimal stop-loss, has been
improved. We conjecture that by increasing number of layer such
convex combination of variances will be improved.

Under criteria of maximizing of the expected utility, one may {\it
either} determine an optimal reinsurance contract (see Kaluszka \&
Okolewski, 2008, for more details) {\it or} estimate unknown
parameters of an optimal reinsurance contract (see Dickson, 2005
\S 9.2, for more details).

The following example considers criteria of maximizing of a convex
combination of the expected exponential utility of $X_R$ and
$X-X_R$ as an additional appropriate criteria to estimate unknown
parameters of a 2-layer optimal reinsurance contract.

\begin{example}\label{Utility_criteria}
Suppose that random claim $X$ has been distributed according to
one of the distributions given in the first column of Table 2.
Moreover consider the optimal 2-layer contract given in Example
(1).

Similar to Example (1), for the sake of simplicity, we set
$M_1=d_\alpha+ d_1$ and $M_2=2d_\alpha+d_1+d_2.$ Now $M_0$ has
been estimated such that $E(X_R)=E(\max\{X-d_\alpha, 0\}).$ Other
two parameters $d_1$ and $d_2$ are estimated such that the
following convex combination of the expected exponential utilities
of $X_R$ and $X-X_R$ has been minimized.
    \begin{eqnarray}\label{Convex_combination_Utilities}
       U_{h}=\omega E(\exp(-\beta(h(X))))+(1-\omega) E(\exp(-\beta(X-h(X)))).
    \end{eqnarray}
where we set $\omega=0.2$ and $\beta_1=\beta_2=1.$
\end{example}
Table 2 shows estimation for unknown parameters of the optimal
2-layer ${X_R^{2-layer}}.$
\begin{center}
    \tiny{Table 2: Estimation for unknown parameters of the optimal
2-layer contract under minimization $U_{h}$ as an optimal criteria, whenever $\omega=0.2$ and $\alpha=0.1$.\\
        \begin{tabular}[c]{c c c c c c c c c c c }
            \hline
            Random claim distribution& $d_\alpha$ &$M_1$&$M_2$&$E(h^{SL}(X))=E(h^{2-layer}(X))$ & $CTE_{h^{SL}}=CTE_{h^{2-layer}}$&$U_{h^{SL}}$&$U_{h^{2-layer}}$\\
            \hline
            Exp(10)       &23.0259 &24.4259&48.4518&$1$& 10.423&$0.9312$&$0.9163$\\
            Exp(8)        &18.4206&31.4132&51.1488&$0.4498$& 8.14&$0.6412$&$0.5629$\\
            Exp(4)        &9.2103&13.2103&23.4037&$0.4099$& 4.0743&$0.8449$&$0.2000$\\
            Weibull(1,2)  &1.5174 &4.1396&6.657&$0.028$&  0.2865&$0.5629$&$0.4593$\\
            Weibull(3,2)  &4.5523 &12.7469&18.2992&$0.02135$& 1.2235&$0.3069$&$0.1465$\\
            \hline
        \end{tabular}}\\
        $Q_{h}$ and $h^*$ are given by Equation \eqref{Convex_combination_Utilities}, $h^{SL}(X)=\max\{X-d_\alpha,0\}$ and $h^{2-layer}(X)=X_R^{2-layer}(X).$ \\
        \normalsize
    \end{center}
The last two columns of Table 2 show the convex combination of
expected exponential utility of $X_R=h(X)$ and $X_I=X-h(X)$ for
the optimal stop-loss and the optimal 2-layer contracts,
respectively. As one may observe, under the optimal 2-layer
contract such convex combination of utilities, compare to optimal
stop-loss contract, is improved.

The Bayesian method under name of the credibility method is
well-known in various areas of the actuarial sciences. For
instance see: Whitney (1918) and Payandeh Najafabdi et al. (2015)
for its application in the experience rating system; Bailey
(1950), Payandeh Najafabdi (2010), and Payandeh Najafabdi et al.
(2012) for its application in evaluating insurance premium;
Hesselager \& witting (1998) and England \& Verral (2002) for its
application in the IBNR claims reserving system; and see Makov et
al. (1996), Makov (2001), and Hossack et al. (1999) for its
general applications in actuarial science.

Now we employ the Bayesian estimation method as an appropriate
method to estimate unknown parameters of an optimal multi-layer
reinsurance contract.

To derive any Bayes estimator for $M^*_0,\cdots, M^*_{m-2},$ based
upon i.i.d. random claim $X^{(1)},\cdots,X^{(n)}.$ One has to
consider initial values for $M^*_0,\cdots,M^*_{m-2}.$ Then, using
such initial values, he/she can define i.i.d reinsurer's random
claim $X^{(1)}_R,\cdots,X^{(n)}_R.$ Now, using information given
by $X^{(1)}_R,\cdots,X^{(n)}_R$ accompanied with prior information
on parameters $M^*_0,\cdots,M^*_{m-2}$ and other unknown
parameters, the Bayes estimator for parameters
$M^*_0,\cdots,M^*_{m-2},$ say $\hat M^*_0,\cdots,\hat M^*_{m-2}$,
under an appropriate loss function can be obtained. Certainly,
such Bayes estimator may be, iteratively, improved by using $\hat
M^*_0,\cdots,\hat M^*_{m-2}$ as a new initial estimator for
$M^*_0,\cdots,M^*_{m-2}.$ And determining
$X^{(1)}_R,\cdots,X^{(n)}_R,$ and finally reevaluating the Bayes
estimator $\hat M^*_0,\cdots,\hat M^*_{m-2},$ again.

Suppose $X^{(1)},\cdots,X^{(n)},$ given parameter $\theta,$ are
i.i.d. random claims with a common density function $f_X$ and a
distribution function $F_X.$ Moreover, suppose that $
m^*_0,\cdots,m^*_{k-2}$ stand for the initial values for $
M^*_0,\cdots,M^*_{k-2}$. Using a straightforward calculation, the
density function for random variable $X_R^{(i)},$ for
$i=1,\cdots,n,$ given parameters $\Theta:=(\theta, M^*_0,\cdots,
M^*_{k-2})$ at observed value $y^{(i)},$ is equal to
{\footnotesize\begin{eqnarray*}
  {g_{{X_R^{(i)}}|\Theta}}({y^{(i)}})  &=&
  \left({F_X}(M^*_0)-{F_X}(0)\right)I_{\{0\}}(y^{(i)})+{f_X}(y^{(i)}+M^*_0)I_{(0,
  M^*_1-M^*_0)}(y^{(i)})\\ &&+\left({F_X}(M^*_2) -
  {F_X}(M^*_1)\right)I_{\{M^*_1-M^*_0\}}(y^{(i)})\\&&+f_X(y^{(i)}+M^*_2- (M^*_1-M^*_0))I_{(M^*_1-M^*_0, M^*_3-M^*_2+
  M^*_1-M^*_0)}(y^{(i)})\\&&+\left({F_X}(M^*_4) -
  {F_X}(M^*_3)\right)I_{\{M^*_3-M^*_2+M^*_1-M^*_0\}}(y^{(i)})+\cdots\\&&+{f_X}\left({y^{(i)}}+M^*_{k-2}-\sum\limits_{j =
                1}^{k/2-2}{({M^*_{2j+1}} - {M^*_{2j
                    }})}- (M^*_1-M^*_0)\right)I_{(\sum\limits_{j = 1}^{k/2-2}{({M^*_{2j+1}} - {M^*_{2j}})}+ (M^*_1-M^*_0),\infty)}(y^{(i)})
\end{eqnarray*}\normalsize}
Using the fact that random variables $X^{(1)}_R,\cdots,X^{(n)}_R$
are i.i.d. Therefore, the joint density function for
$X^{(1)}_R,\cdots,X^{(n)}_R,$ given parameters $\Theta:=(\theta,
M^*_0,\cdots, M^*_{k-2})$  can be restated as
{\footnotesize\begin{eqnarray*}
  f_{X^{(1)}_R,\cdots,X^{(n)}_R}(y^{(1)},\cdots,y^{(n)}|\Theta) &=& {[{F_X}(M^*_0)-{F_X}(0)]^{n_0}}\prod\limits_{i = 1}^{{n_1}} {{f_X}(y^{(i)}+M^*_0)} {[{F_X}(M^*_2) - {F_X}(M^*_1)]^{{n_2}}}
  \cdots\\ &&\times\prod\limits_{i = {n_0} +  \cdots {n_{(k - 2)}}}^n
            {{f_X}\left({y^{(i)}}+M^*_{k-2}-\sum\limits_{i =
                    1}^{k/2-2}{({M^*_{2i+1}} - {M^*_{2i }})}-
                    (M^*_1-M^*_0)\right)},
\end{eqnarray*}\normalsize}
where $n_{0}:=\#(y^{(i)}=0),$ $n_{1}:=\#(0<y^{(i)}<
(M^*_1-M^*_0)),$ $n_{2}:=\#(y^{(i)}= (M^*_1-M^*_0)),\cdots,
n_{k-2}:=\#(\sum\limits_{i = 1}^{k/2-2}{({M^*_{2i+1}} -{M^*_{2i
}})} <y^{(i)})$.

Assuming $\pi {(\theta,M^*_0,...,M^*_{k-2})}$ is the prior
distribution for vector ${(\theta,M^*_0,\cdots,M^*_{m-2})}$, joint
posterior distribution for vector
$\Theta:=(\theta,M^*_0,\cdots,M^*_{k-2})$ is {\footnotesize
\begin{eqnarray*}
            &&\pi(\theta,M^*_0,\cdots,M^*_{m-2}|y^{(1)},\cdots,y^{(n)})
            =\\
            &&\frac{f_{X^{(1)}_R,\cdots,X^{(n)}_R}(y^{(1)},\cdots,y^{(n)}|\theta,M^*_0,\cdots,M^*_{k-2})
                \pi (\theta,M^*_0,\cdots,M^*_{k-2})}{\int_{\mathcal{M^*}_{k-2}}\cdots
                \int_\Theta
                f_{X^{(1)}_R,\cdots,X^{(n)}_R}(y^{(1)},\cdots,y^{(n)}|\theta,M^*_0,\cdots,M^*_(k-2))
                \pi (\theta,M^*_0,\cdots,M^*_{k-2})d\theta dM^*_0,\cdots,dM^*_{k-2}}.
        \end{eqnarray*}\normalsize}
Using the above joint posterior distribution, the Bayes estimator
for each $M^*_0,...,M^*_{k-2}$ under the square error loss
function, is
        \begin{eqnarray}
            \label{Bayes-for-M_s}
            {\hat M^*}_{i} &=& \int_{\mathcal{M^*}_{k-2}}\cdots \int_\Theta M^*_i~\pi (\theta,M^*_0,\cdots,M^*_{k-2}|y^{(1)},\cdots,y^{(n)})d\theta
            dM^*_0 \cdots dM^*_{k-2},
        \end{eqnarray}
for $i=0,\cdots,k-2.$

Now as an application of the above findings, we consider the
following example.
\begin{example}\label{Bayesian_criteria2}
Suppose that random claim $X$ has been distributed according to
one of the distributions given in the first column of Table 3.
Moreover, suppose that the optimal multi-layer contract has 1
layer and restated as
\begin{eqnarray*}
        {X_R^{1-layer}} = \left\{ \begin{array}{l}
            0\,\,\,\,\,\,\,\,\,\,\,\,\,\,\,\,\,\,\,\,\,\,\,\,\,\,\,\,\,\,\,\,\,\,\,\,\,\,\,\,\,\,\,\,\,\,\,\,\,\,\,\,\,\,\,\,\,\,\,\,\,\,\,\,\,\,\,\,\,\,\,\,\,\,\,\,\,\,\,\,\,\,\,\,\,\,\,\,\,\,\,\,\,\,\,\,\,\,\,\,\,\,\,\,\,\,\,\,\,\,\,X < {M_0}\\
            X - {M_0}\,\,\,\,\,\,\,\,\,\,\,\,\,\,\,\,\,\,\,\,\,\,\,\,\,\,\,\,\,\,\,\,\,\,\,\,\,\,\,\,\,\,\,\,\,\,\,\,\,\,\,\,\,\,\,\,\,\,\,\,\,\,\,\,\,\,\,\,\,\,\,\,\,\,\,\,\,{M_0} \le X < {M_1}\\
            {M_1} - {M_0}\,\,\,\,\,\,\,\,\,\,\,\,\,\,\,\,\,\,\,\,\,\,\,\,\,\,\,\,\,\,\,\,\,\,\,\,\,\,\,\,\,\,\,\,\,\,\,\,\,\,\,\,\,\,\,\,\,\,\,\,\,\,\,\,\,\,\,\,\,\,\,\,\,\,{M_1} \le X < {M_2}\\
            X + ({M_1} - {M_0}) -
            {M_2}\,\,\,\,\,\,\,\,\,\,\,\,\,\,\,\,\,\,\,\,\,\,\,\,\,\,\,\,\,\,\,\,\,\,\,\,\,\,\,\,\,{M_2}
            \le X
        \end{array} \right..
\end{eqnarray*}
For the sake of simplicity, we set $d_0=M_0,$ $d_1=M_1-M_0,$ and
$d_2=M_2-M_1.$ Now, suppose that the prior distributions of the
unknown parameters $d_0,$ $d_1,$ and $d_2$ are independent and
given in the second, third, and fourth columns of Table 3,
respectively.
\end{example}
To construct a Bayes estimator for unknown parameters, we employed
$d_0=0.20,$ $d_1=0.15,$ and $d_2=0.02$ as initial values.

\begin{landscape}
    \begin{scriptsize}
        \begin{center}
            \tiny Table 3: Mean (standard deviation) of Bayes estimator for$d_0$, $d_1$ and
$d_2$ based upon 100 sample size and 100 iterations, whenever $\alpha=0.1$.\\
            \begin{tabular}[c]{c c c c c c c c c c }
                \hline\\
                Claim Distribution& Prior distribution& Prior distribution&Prior distribution& Mean (variance)  & Mean (variance)  & Mean (variance) & $E(h^{SL}(X))=$ & & $CTE_{h^{SL}}=$\\
                &for $d_0$&for $d_1$&for $d_2$& of estimated $d_0$ &  of
                estimated $d_1$ & of
                estimated $d_2$ & $E(h^{1-layer}(X))$ & $h^{SL}(X)$ & $CTE_{h^{1-layer}}$\\\hline \\
                EXP(1)  &EXP(1)      &EXP(1)       &EXP(1)       &0.0599&0.4474& 0.0643   & 0.1 & $(X-2.3026)_+$ & 1.01   \\
                &&                                                        &             &(4.795$\times 10^{-16}$)&(4.439$\times 10^{-14}$)&(8.458$\times 10^{-7}$)\\ \\
                EXP(4)  &Gamma(2,3)      &Gamma(3,2)       &Gamma(2,2)&0.0526&0.6575&0.0638 & 0.4 & $(X-9.2103)_+$ & 4.0743  \\
                &&                                                                                     &&(3.823$\times 10^{-18}$)&(9.003$\times 10^{-13}$)&(1.093$\times 10^{-5}$)\\ \\
                Weibull(1,2)    &Gamma(2,2)      &Gamma(3,2)& Gamma(2,3)      &0.0746&0.6575&0.0798 & $0.028249$ & $(X-1.5174)_+$ & 0.2865 \\
                &&                                                          &                           &(1.661$\times 10^{-17}$)&(4.393$\times 10^{-15}$)&(3.542$\times 10^{-6}$)\\ \\
                \hline
            \end{tabular}\\
            $h^{SL}(X)=\max\{X-d_\alpha,0\}$ and $h^{1-layer}(X)=X_R^{1-layer}(X).$\\
            \normalsize
        \end{center}
    \end{scriptsize}
\end{landscape}

The three last columns of Table 3 represent the mean and the
standard deviation, respectively, of the Bayes estimator for
$d_0$, $d_1$ and $d_2$, which generates 100 random numbers from a
given distribution. This estimators were derived using Equation
\eqref{Bayes-for-M_s} when the mean of 100 iterations of the Bayes
estimator for $d_0$, $d_1$ and $d_2$ was used as an estimator for
$d_0$, $d_1$ and $d_2.$

The small variance of these estimators shows that the estimation
method is an appropriate method to use with the different samples.
\section{Conclusion and suggestions}
This article generalizes the stop-loss reinsurance policy to a new
continuous multi-layer reinsurance policy which minimizes the
conditional tail expectation (CTE) risk measure of the insurer's
total risk. Unknown parameters of the new optimal multi-layer
reinsurance policy can be estimated using other additional
appropriate criteria. Therefore, the new multi-layer reinsurance
policy {\it not only} similar to the original stop-loss
reinsurance policy is optimal, in a same sense, {\it but also} it
has some other appropriate criteria which the original stop-loss
policy does not have. Estimation method of this article can be
generalized to the other appropriate criteria such as the ruin
probability (Fang \& Qu, 2014), percentile matching estimating
method (Teugels \& Sundt, 2004), etc.

The following two propositions are generalized result of this
article under the general translative and monotone risk measure
$\rho(\cdot)$.

The following suppose that under minimization criteria of a
translative and monotone risk measure $\rho(\cdot)$ of the
insurer's total risk reinsurance contract $f(\cdot)$ is optimal.
Then, it provides a multi-layer reinsurance contract which its
corresponding risk measure coincides with the insurer's total risk
under contract $f(\cdot),$  see Figure 2(a) for an illustration.
\begin{proposition}
\label{common-risk-measure} Suppose $\rho(\cdot)$ is a translative
and monotone risk measure. Moreover, suppose that $f(\cdot)$ in
the class of reinsurance strategies $\mathcal{C}$ minimizes risk
measure of the total risk of insurance company. Then, reinsurance
$g(\cdot)$ also minimizes the risk measure of total risk of
insurance company.
\begin{eqnarray*}
  g(X) &=&
  f(X)I_{[0,M_1)}(X)+(X-M_1+f(M_1))I_{[M_1,M_2)}(X)+f(M_2^*)I_{[M_2,M_2^*)}(X)\\&&+(X-M_2^*+f(M_2^*))I_{[M_2^*,M_3)}(X)+\cdots+(X-M_k^*+f(M_k^*))I_{[M_k^*,\infty)}(X),
\end{eqnarray*}
where $M_1,M_2,\cdots,M_k$ are unknown parameters of the new
optimal reinsurance and $M_1^*,M_2^*,\cdots,M_k^*$ have to be
evaluated using equation $f(M_2^*)=M_2-M_{1}+f(M_1)$ and
$f(M_i^*)=M_i-M_{i-1}^*+f(M_{i-1}^*)$ for $i=3,\cdots,k.$
\end{proposition}
{\it Proof.} Since $\rho(\cdot)$ is a translative risk measure,
one may write that
\begin{eqnarray*}
\rho(X-g(X)+\pi_g^X)&=&\rho(X-g(X))+\pi_g^X\\
                    &=& \rho\left[(X-f(X))I_{[0,M_1)}(X)+(M_1-f(M_1))I_{[M_1,M_2)}(X)\right.\\
                    && ~~~~\left.+(X-f(M_2^*)I_{[M_2,M_2^*)}(X))+(M_2^*-f(M_2^*))I_{[M_2^*, M_3)}(X)\right.\\
                    && ~~~~\left.+(X-f(M_3^*))I_{[M_3, M_3^*)}+\cdots+(M_k^*-f(M_k^*))I_{[M_k^*,\infty)}(X)\right]+\pi_g^X\\
                    &\leq & \rho(X-f(X))+\pi_g^X\\
                    &=&\rho(X-f(X)+\pi_g^X)\\
                    &=&\rho(X-f(X)+\pi_f^X).
\end{eqnarray*}
The above inequality arrives from the fact that $\rho(\cdot)$  is
a monotone risk measure and $X-g(X)\leq X-f(X)$ with probability
1. Now using the fact that $\rho(X-f(X))= \mathop {\min
}\limits_{h \in \mathcal{C}}\rho(X-h(X)+\pi_h^X)$ we conclude that
the above inequality has to be changed to an equality. $\square$

Now we provide an optimal multi-layer reinsurance contract, for a
situation that the optimal reinsurance $f(\cdot)$ arrives by
minimizing a convex combination of two translative and monotone
risk measures $\rho_1(\cdot)$ and $\rho_2(\cdot)$ of the insurer's
total risk, $X_R=h(X),$ and the reinsurer's total risk
$X_I=X-h(X),$ i.e.,
$f(X)=\operatornamewithlimits{argmin}_{h\in\mathcal{C}}\{\omega\rho_1(X-h(X)+\pi_h^X)+(1-\omega)\rho_2(h(X)-\pi_h^X)\},$
where $\omega\in[0,1],$ see Figure 2(b) for an illustration.

As an example for such optimal reinsurance $f(\cdot),$ under such
the convex combination of two distortion risk measures, see Assa
(2015).
\begin{proposition}
\label{Convex_combination_general-risk-measures} Suppose
$\rho_1(\cdot)$ and $\rho_2(\cdot)$ are two translative and
monotone risk measures. Moreover, suppose that $f(\cdot)$ in class
of reinsurance strategies $\mathcal{C}$ minimizes a convex
combination of two risk measures $\rho_1(\cdot)$ and
$\rho_2(\cdot),$ i.e.,
$f(X)=\operatornamewithlimits{argmin}_{h\in\mathcal{C}}\{\omega\rho_1(X-h(X)+\pi_h^X)+(1-\omega)\rho_2(h(X)-\pi_h^X)\},$
where $\omega\in[0,1].$ Then, for
$\omega^*\in(0,a_{\min}/(a_{\min}+a_{\max})),$ the following
k-layer reinsurance $g(\cdot)$ also minimizes such the convex
combination of two risk measures $\rho_1(\cdot)$ and
$\rho_2(\cdot)$.
\begin{eqnarray*}
  g(X) &=&
  f(X)I_{[0,M_1)}(X)+(X-M_1+f(M_1))I_{[M_1,M_2)}(X)+f(M_2^*)I_{[M_2,M_3)}(X)\\&&+(X-M_3+f(M_2^*))I_{[M_3,M_4)}(X)+\cdots+f(X)I_{[M_{2k+1}^*,\infty)}(X),
\end{eqnarray*}
where $M_1,M_2,\cdots,M_k$ are unknown parameters of the new
optimal reinsurance and $M_1^*,M_2^*,\cdots,M_k^*$ have to be
evaluated using equation $f(M_2^*)=M_2-M_{2}+f(M_1),$
$f(M_{2j-1}^*)=M_{2j-1}^*-M_{2j-1}+f(M_{2(j-1)}^*),$
$f(M^*_{2j})=f(M^*_{2(j-1)})+M_{2j}-M_{2j-1},$ for $j=2,\cdots,k,$
$a_{\min}:=\operatornamewithlimits{min}_{x\in A}\{|2f(x)-x|\},$
$a_{\max}:=\operatornamewithlimits{max}_{x\in A}\{|2f(x)-x|\}$ and
$A:=[M^1,M^*_{2})\cup_{j=2}^{k}[M^*_{2j-1},M^*_{2j}].$
\end{proposition}
{\it Proof.} Set $\pi_g^*:=\omega^*\pi_g^X-(1-\omega^*)\pi_g^X.$
Since $\rho_1(\cdot)$ and $\rho_2(\cdot)$ are a translative risk
measures, one may write that{\footnotesize
\begin{eqnarray*}
\omega^*\rho_1(X-g(X)+\pi_g^X)+(1-\omega^*)\rho_2(g(X)-\pi_g^X)&=&\pi_g^*+\omega^*\rho_1(X-g(X))+(1-\omega^*)\rho_2(g(X)\\
                    &\leq&\pi_g^*+ \omega^*\rho_1\left[(X-f(X))I_{[0,M_1)}(X)+f(X)I_{[M_1,M_2^*)}(X)\right.\\
                    &&~~~~~~~~~~\left.+(X-f(X))I_{[M_2^*, M_3^*)}(X)+\cdots+(X-f(X))I_{[M_{2k+1}^*,\infty)}(X)\right]\\
                    &&+(1-\omega^*)\rho_2\left[f(X)I_{[0,M_1)}(X)+(X-f(X))I_{[M_1,M_2^*)}(X)\right.\\
                    && ~~~~~~~~~~~~~~~~\left.+f(X)I_{[M_2^*, M_3^*)}(X)+\cdots+f(X)I_{[M_{2k+1}^*,\infty)}(X)\right]\\
                    &=&\pi_g^*+\omega^*\rho_1\left[(X-f(X))I_{[0,\infty)}(X)+(2f(X)-X)I_{[M^1,M^*_{2})}(X))\right.\\
                    &&~~~~~~~~~~\left.+(2f(X)-X)\sum_{j=2}^{k}I_{[M^*_{2j-1},M^*_{2j})}(X))\right.]\\
                    &&+(1-\omega^*)\rho_2\left[f(X)I_{[0,\infty)}(X)+(X-2f(X))I_{[M^1,M^*_{2})}(X))\right.\\
                    && ~~~~~~~~~~~~~~~~\left.+(X-2f(X))\sum_{j=2}^{k}I_{[M^*_{2j-1},M^*_{2j})}(X))\right.]\\
                    &\leq & \pi_g^*+ \omega^*\rho_1\left[(X-f(X))I_{[0,\infty)}(X)\right]+(1-\omega^*)\rho_2\left[f(X)\right]\\
                    &&+\omega^*ka_{\max}-(1-\omega^*)ka_{\min}\\
                    &\leq & \pi_g^*+ \omega^*\rho_1\left[(X-f(X))I_{[0,\infty)}(X)\right]+(1-\omega^*)\rho_2\left[f(X)\right]\\
                    &=&\omega^*\rho_1(X-f(X)+\pi_g^X)+(1-\omega^*)\rho_2(f(X)-\pi_g^X)\\
                    &=&\omega^*\rho_1(X-f(X)+\pi_f^X)+(1-\omega^*)\rho_2(f(X)-\pi_f^X).
\end{eqnarray*}}\normalsize
The last inequality arrives from the fact that
$\omega^*\in[0,a_{\min}/(a_{\min}+a_{\max}).$ Now using the fact
that
$\omega^*\rho_1(X-f(X)+\pi_f^X)+(1-\omega^*)\rho_2(f(X)-\pi_f^X)=
\mathop {\min }\limits_{h \in
\mathcal{C}}\left\{\omega^*\rho_1(X-h(X)+\pi_h^X)+(1-\omega^*)\rho_2(h(X)-\pi_h^X)\right\},$
we conclude that the k-layer reinsurance $g(\cdot)$ also minimizes
such the convex combination. $\square$

\begin{center}
\begin{figure}[h!]
\centering\subfigure[]{
\includegraphics[width=7cm,height=6cm]{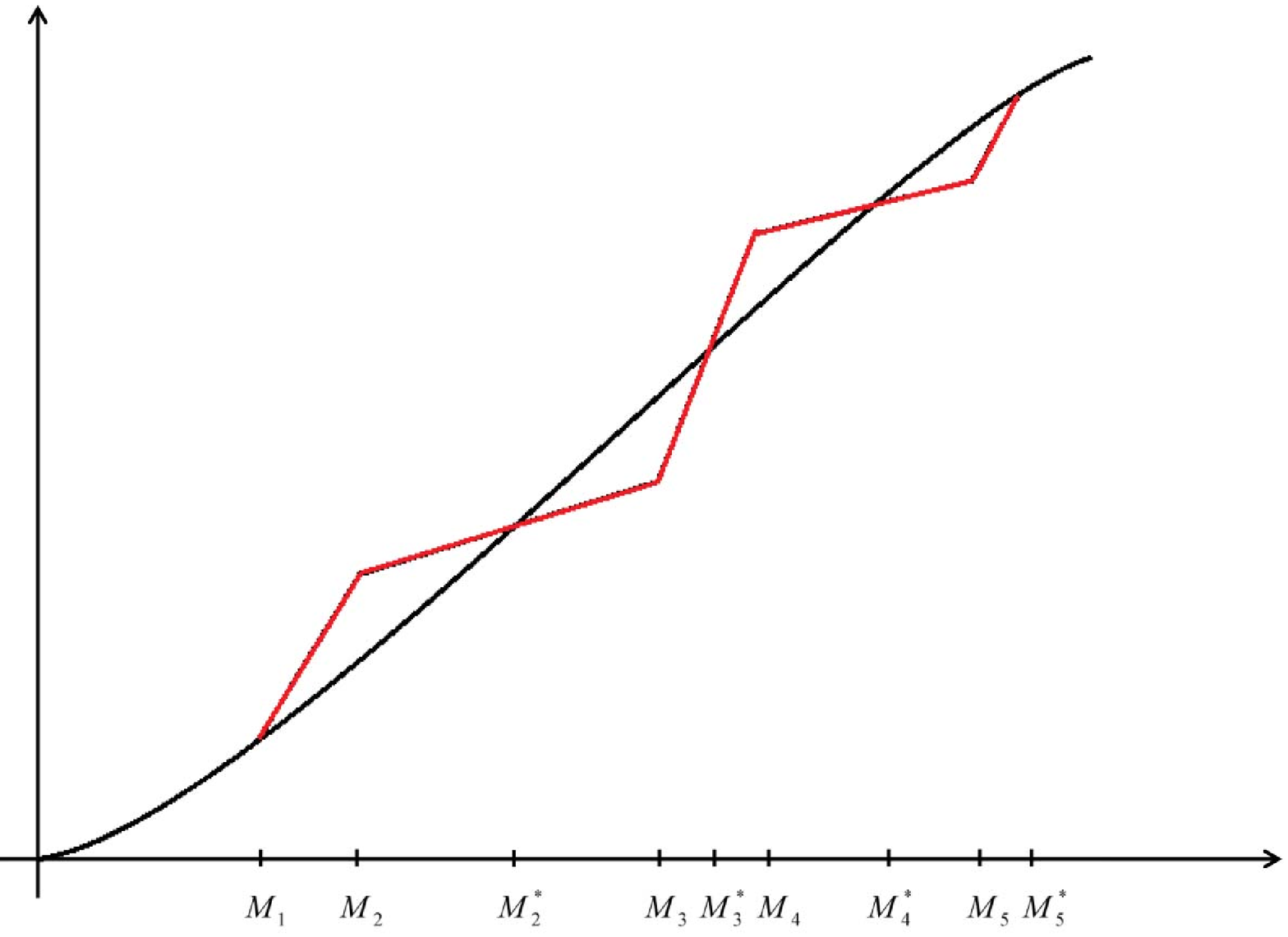}}\subfigure[]{
\includegraphics[width=7cm,height=7cm]{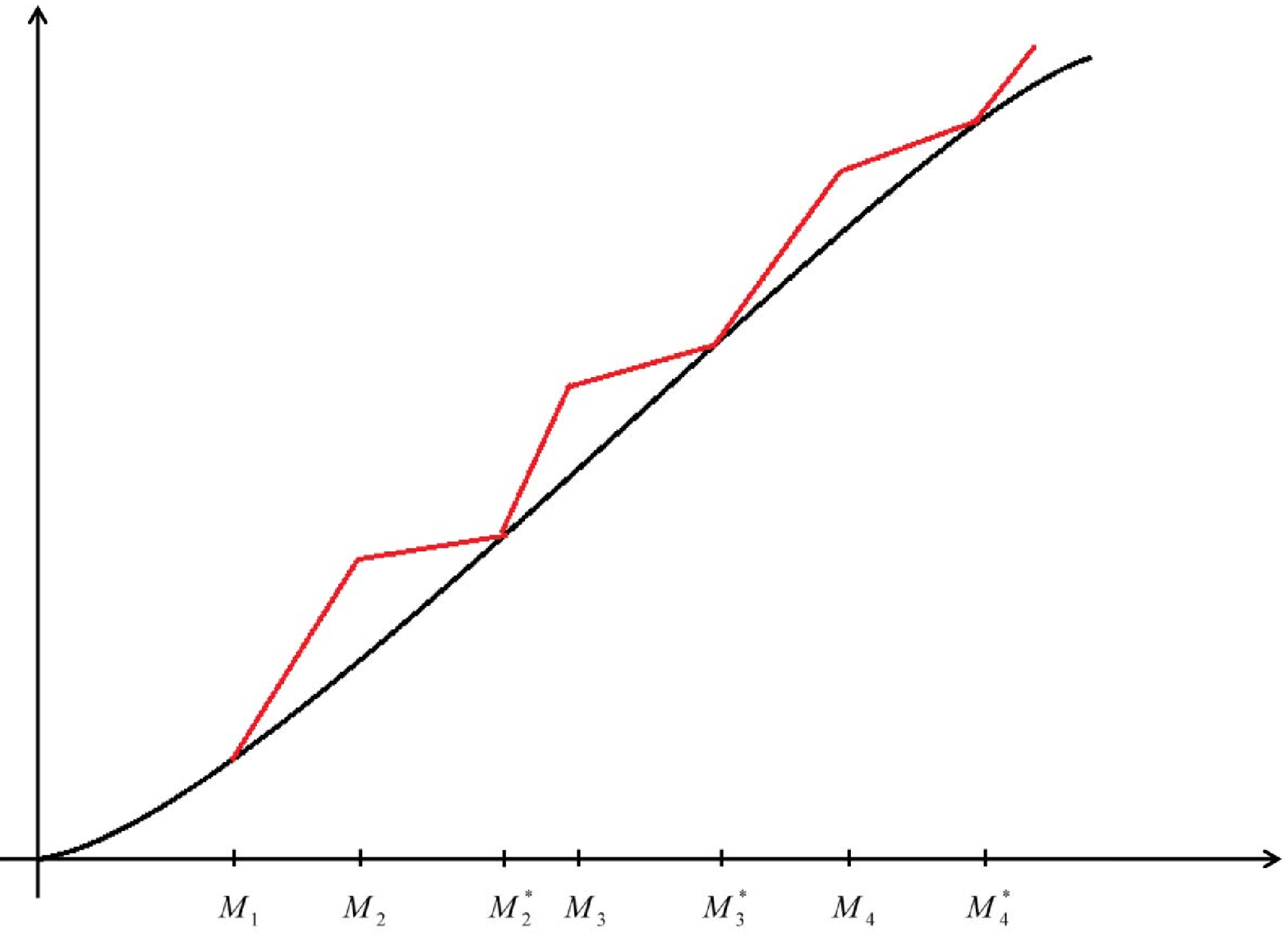}
} \caption{\scriptsize Part (a): The optimal multi-layer
reinsurance contract, given by Proposition
\eqref{common-risk-measure} whenever
$f(X)=\operatornamewithlimits{argmin}_{h\in\mathcal{C}}\{\rho(X-h(X)+\pi_h^X)\}$
and Part (b): The optimal multi-layer reinsurance contract, given
by Proposition \eqref{Convex_combination_general-risk-measures},
whenever
$f(X)=\operatornamewithlimits{argmin}_{h\in\mathcal{C}}\{\omega\rho_1(X-h(X)+\pi_h^X)+(1-\omega)\rho_2(h(X)-\pi_h^X)\}$
and $\omega\in[0,1].$}
\end{figure}
\end{center}
\section{Acknowledgements}
The second author also would like thanks support of the Central
Insurance of the Islamic Republic of Iran.


\begin{thebibliography}{000}
\bibitem{Assa 2015} Assa, H. (2015). On Optimal Reinsurance Policy with Distortion Risk Measures and Premiums. {\it Insurance: Mathematics and Economics}, {\bf 61}, 70--75.
\bibitem{Bailey 1950} Bailey, A.L. (1950). Credibility Procedures Laplace's Generalization of Bayes' Rule and the Combination of Collateral Knowledge with Observed Data. {\it Proceedings of the Casualty Actuarial Society} {\bf 37}, 7--23.
\bibitem{Borch 1960} Borch, K. (1960). An Attempt to Determine the Optimum Amount of Stop-Loss Reinsurance. in Transactions of the 16th International Congress of Actuaries, 597--610.
\bibitem{Chi2012a} Chi, Y. (2012). Reinsurance Arrangements Minimizing the Risk-Adjusted Value of an Insurer's Liability. {\it Astin Bulletin}, {\bf 42}(02), 529--557.
\bibitem{Chi2012b} Chi, Y. (2012). Optimal reinsurance under variance related premium principles. {\it Insurance: Mathematics and Economics}, {\bf 51}(2), 310--321.
\bibitem{Cai 2012} Cai, J., Fang, Y., Li, Z., \& Willmot, G. E. (2013). Optimal Reciprocal Reinsurance Treaties under the Joint Survival Probability and the Joint Profitable Probability. {\it Journal of Risk and Insurance}, {\bf 80}, 145--168.
\bibitem{Cai 2011} Chi, Y., \& Tan, K. S. (2011). Optimal Reinsurance under VaR and CVaR Risk Measures: a Simplified Approach. {\it ASTIN Bulletin}, {\bf 41}, 487--509.
\bibitem{Cai 2008} Cai, J., Tan, K. S., Weng, C., \& Zhang, Y. (2008). Optimal Reinsurance under VaR and CTE Risk Measures. {\it Insurance: Mathematics and Economics}, {\bf 43}, 185--196.
\bibitem{Chi-Tan2013} Chi, Y., \& Tan, K. S. (2013). Optimal reinsurance with general premium principles. {\it Insurance: Mathematics and Economics}, {\bf 52}(2), 180--189.
\bibitem{Cai-Weng2014} Cai, J., \& Weng, C. (2014). Optimal reinsurance with Expectile. {\it Scandinavian Actuarial Journal}, 1--22.
\bibitem{Cortes2013} Cortes, O. A. C., Rau-Chaplin, A., Wilson, D., Cook, I., \& Gaiser-Porter, J. (2013). Efficient optimization of reinsurance contracts using discretized PBIL. Data Analytics: The Second International Conference on Data Analytics. Porto, Portuga
\bibitem{Dedu 2012} Dedu, S. (2012). Optimization of some Risk Measures in Stop-Loss Reinsurance with Multiple Retention Levels. {\it Mathematical Reports}, {\bf 14}, 131--139.
\bibitem{Denuit-et-al} Denuit, M., Dhaene, J., Goovaerts, M., \& Kaas, R. (2006). {\it Actuarial Theory for Dependent Risks: Measures, Orders and Models}. John Wiley \& Sons.
\bibitem{Dickson} Dickson, D. C. (2005). {\it Insurance risk and ruin.} Cambridge University Press, New York.
\bibitem{Fang-Qu} Fang, Y., \& Qu, Z. (2014). Optimal Combination of Quota-Share and Stop-Loss Reinsurance Treaties under the Joint Survival Probability. {\it IMA Journal of Management Mathematics}, {\bf 25}, 89--103.
\bibitem{England 2002} England, P. \& Verrall, R. (2002). Stochastic Claims Reserving in General Insurance (with discussion). {\it British Actuarial Journal}. {\bf 8}, 443--544.
\bibitem{Hesselager 1990} Hesselager, O. (1990). Some Results on Optimal Reinsurance in Terms of the Adjustment Coefficient. {\it Scandinavian Actuarial Journal}. {\bf 1990}, 80--95.
\bibitem{Hesselager 1988}  Hesselager, O. \& Witting, T. (1988). A Credibility Model with Random Fluctuations in Delay Probabilities for the Prediction of IBNR Claims. {\it ASTIN Bulletin}, {\bf 18}, 79--90.
\bibitem{Hossack 1999} Hossack, I.B., Pollard, J.H. \& Zenwirth, B. (1999). {\it Introductory Statistics with Applications in General Insurance}, 2nd Edition, University Press, Cambridge.
\bibitem{Kaluszka} Kaluszka, M. (2005). Truncated stop loss as optimal reinsurance agreement in one-period models. {\it Astin Bulletin}, {\bf 35}(02), 337--349.
\bibitem{Kaluszka-Okolewski} Kaluszka, M., \& Okolewski, A. (2008). An extension of Arrow's result on optimal reinsurance contract. {\it Journal of Risk and Insurance}, {\bf 75}(2), 275--288.
\bibitem{Makov 2001} Makov, U. E. (2001). Principal Applications of Bayesian Methods in Actuarial Science: a Perspective. {\it North American Actuarial Journal} {\bf 5}, 53--73.
\bibitem{Makov 1996} Makov, U. E., Smith, A.F.M. \& Liu, Y.H. (1996). Bayesian Methods in Actuarial Science. {\it The Statistician}, {\bf 45}, 503--515.
\bibitem{Ouyang-li} Ouyang, Y. X., \& Li, Z. Y. (2010). Adverse selection, systematic risks and sustainable development of Policy Agricultural Insurance. {\it Insurance Studies}, {\bf 4}, 1--9.
\bibitem{Panahi-Payandeh} Panahi Bazaz, A. \& Payandeh Najafabadi, A. T. (2015). An Optimal Reinsurance Contract from Insurer's and Reinsurer's Viewpoints. {\it Applications \& Applied Mathematics}, {\bf 10}(2), 970--982.
\bibitem{Passalacqua} Passalacqua, L. (2007). Measuring effects of excess-of-loss reinsurance on credit insurance risk capital. {\it Giornale del\l'Istituto Italiano degli Attuari}, {\bf LXX}, 81-102.
\bibitem{Payandeh2010} Payandeh Najafabadi, A. T. (2010). A New Approach to the Credibility Formula. {\it Insurance: Mathematics and Economics}, {\bf 46}, 334--338.
\bibitem{Payandeh2012} Payandeh Najafabadi, A. T., Hatami, H., \& Omidi Najafabadi, M. (2012). A Maximum Entropy Approach to the Linear Credibility Formula. {\it Insurance: Mathematics and Economics}, {\bf 51}, 216--221.
\bibitem{Payandeh4} Payandeh Najafabadi, A. T., \& Qazvini, M. (2015). A GLM Approach to Estimating Copula Models. {\it Communications in Statistics-Simulation and Computation}, {\bf 44}(6), 1641--1656.
\bibitem{Payandeh2016} Payandeh Najafabadi, A. T. \& Panahi Bazaz, A. P. (2016). An optimal co-reinsurance strategy. {\it Insurance: Mathematics and Economics}, {\bf 69}, 149--155.
\bibitem{Porth-2013} Porth, L., Seng Tan, K., \& Weng, C. (2013). Optimal reinsurance analysis from a crop insurer's perspective. {\it Agricultural Finance Review}, {\bf 73}(2), 310--328.
\bibitem{Tan 2010} Tan, K. S. \& Weng, C., (2012). Enhancing Insurer Value using Reinsurance and Value-at-Risk Criterion. {\it The Geneva Risk and Insurance Review}, {\bf 37}, 109--140.
\bibitem{Tan 2011} Tan, K. S., Weng, C., \& Zhang, Y. (2011). Optimality of General Reinsurance Contracts under CTE Risk Measure. {\it Insurance: Mathematics and Economics}, {\bf 49}, 175--187.
\bibitem{Teugels 2004} Teugels, J. L., \& Sundt, B. (2004). {\it  Encyclopedia of Actuarial Science.} Vol. 1. Wiley, New York.
\bibitem{Weng2009} Weng, C. (2009), {\it Optimal reinsurance designs: from an insurer's perspective}. PhD thesis, University of Waterloo, Waterloo, Canada.
\bibitem{Whitney 1918} Whitney, A. W. (1918). The Theory of Experience Rating. {\it Proceedings of the Casualty Actuarial Society} {\bf 4}, 274--292.
\bibitem{Zhuang2016} Zhuang, S. C., Weng, C., Tan, K. S., \& Assa, H. (2016). Marginal Indemnification Function formulation for optimal reinsurance. {\it Insurance: Mathematics and Economics}, {\bf 67}, 65--76.
\end{thebibliography}
\end{document}